# MRADSIM-Converter: A new software for STEP to GDML conversion

Ali Behcet Alpat

*Istituto Nazionale di Fisica Nucleare (INFN) Perugia Section, Via A.Pascoli, 06123, Perugia, Italy*
*(Email: behcet.alpat@pg.infn.it)*

Abdullah Coban

*IRADETS inc., Teknopark Istanbul, Istanbul, Turkiye and Physics Department, Akdeniz University,*
*07058 Antalya, Turkiye*
*(Email: abdullah.coban@iradets.com)*

Hakan Kaya

*IRADETS inc., Teknopark Istanbul, Istanbul, Turkiye*
*(Email: hakan.kaya@iradets.com)*

Giovanni Bartolini

*BEAMIDE srl, Via Campo di Marte 4/O, 06124, Perugia, Italy*
*(Email: giovanni.bartolini@beamide.com)*

**Abstract**

Radiation effects analysis of instruments operative in harsh radiation environment is crucial for performance and functionality of electronic devices and components. Engineering design of instruments is usually carried out in Computer Aided Design (CAD) engineering software. Geant4-based Monte Carlo codes are extensively used for particle transport simulation and analysis. However, Geant4 is not prepared to accept CAD Standard for The Exchange of Product data (STEP) format. MRADSIM-Converter is a new software for STEP to Geometry Description Markup Language (GDML) format conversion, readable by Geant4-based Monte Carlo codes. Its validation with two different converters confirms its higher speed for importing CAD geometries with arbitrary size and complexity having a user-friendly interface for modifying volumes properties.

## I. SOFTWARE DESCRIPTION

Complex geometries are extensively designed and implemented with the application of CAD software available by various companies, where the users find less difficulties modeling their projects for further analysis. However, in case of radiation analysis, geometry implementation is limited due to non-user-friendly structure of codes for Monte Carlo simulations. For instance, Geant4 toolkit simulation code developed at CERN is widely used for high energy physics community as well as health physics and radioprotections communities [8]-[10]. However, Geant4 is not prepared to receive the CAD formats file and the geometry implementation in Geant4 is a time-consuming task as it requires the creation of each volume using specific classes. Also, in case of satellite modeling, thousands of components might be implemented. This fact can increase the possibility of errors in insertion, by-hand, of geometries.

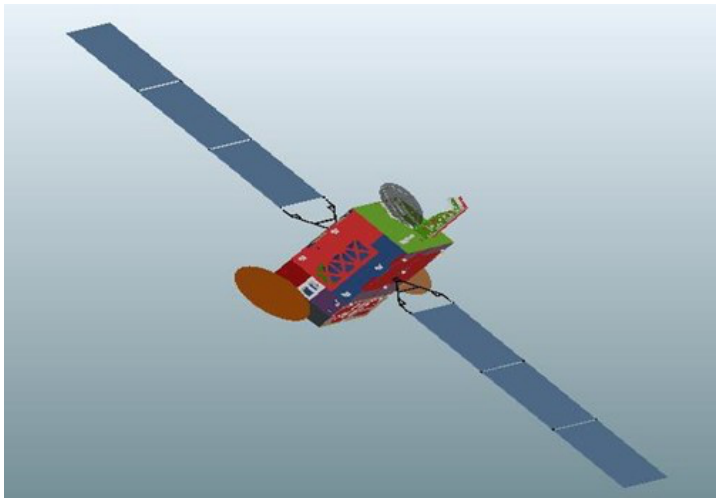

FIG. 1: CAD geometry imported in MRADSIM-Converter

In order to avoid the manual implementation of complex geometries designed inside CAD software into the Geant4 code, a conversion between the files written in STEP format prepared by CAD software into the GDML format, accepted as input in Geant4 is needed. MRADSIM-Converter is a software designed to perform this task. All existent volumes in the input STEP file are visualized in the Graphical User Interface (GUI), where it is possible to interactively assign the material to the components and modify each volume's properties, such as material, colour, and name.

MRADSIM-Converter is tested under Ubuntu, versions 18 and 20. All the libraries necessary to run the program are provided in the package hence no additional library or software is required to be installed. Under OS and Windows, MRADSIM-Converter is running through virtual machine (VM) with above-mentioned Ubuntu versions.

### 1.1. Software architecture and functionalities

The MRADSIM-Converter diagram is shown in Fig. 2. A GDML or STEP format file is imported in the program as an input. The new constructed GDML file can be exported applicable for Geant4-based Monte Carlo codes. A material database is implemented to assign the material for each volume and the CAD processing unit is defined for the STEP file analysis and conversion. All the units are organized and updated through the GUI.

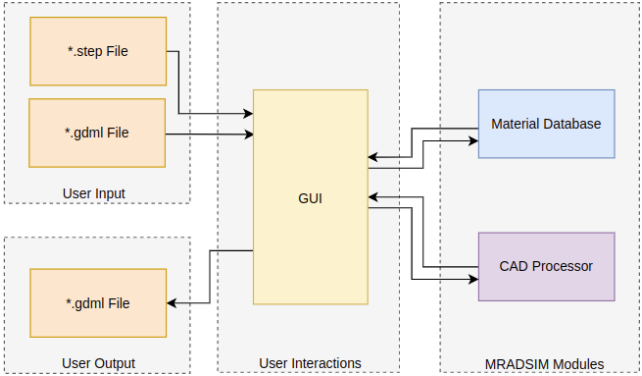

FIG. 2: MRADSIM-Converter diagram

### 1.1.1. GUI

The GUI was developed in C++ using Qt5 framework [11]. Open CASCADE library [12]-[13] is used for 3D graphical interaction capabilities. Fig. 3 shows the MRADSIM-Converter's GUI. It is composed of four panels. All the components available in the geometry of the CAD file are shown in the part list and their dedicated parameters are shown in the information panel, placed on the left. In the center, a 3D view of the geometry is depicted, where the user has the complete control of checking all the components with the implemented options available in the program. The material selecting panel is placed on the right. The user can select a predefined material or build a new one if required.

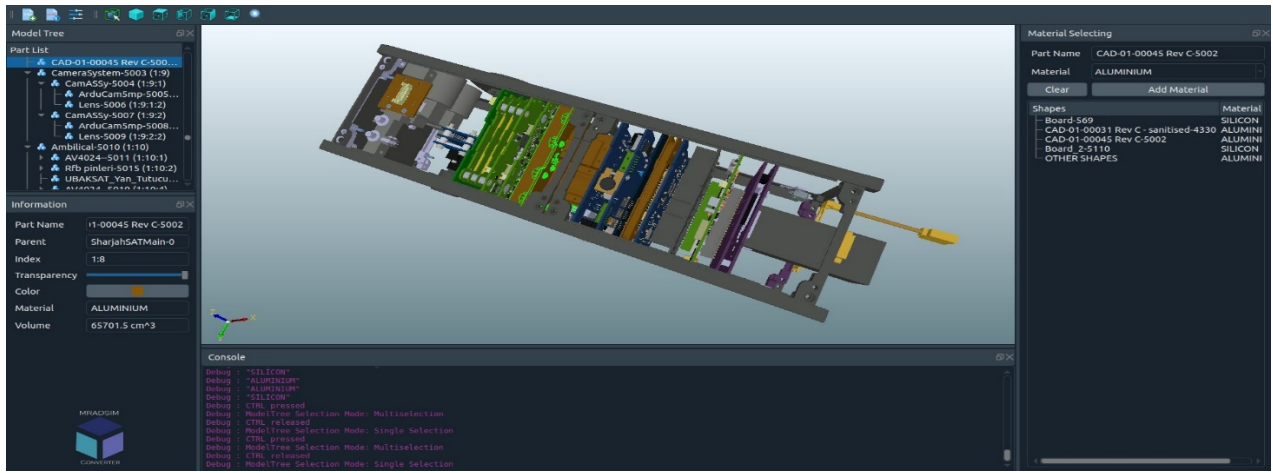

FIG. 3: MRADSIM-Converter GUI. Part list and part information are on the left side. On the right side the material can be selected for each component. The geometry is shown in the center.

### 1.1.2. STEP

STEP format is an ISO standard exchange format with 3D model containing saved data in a text format, easily recognized by CAD programs having high accuracy. It was basically developed to solve the problem of sharing models among users with various CAD programs. However, STEP file cannot contain the material identifiable by CAD software. Also, STEP format is a complex format similar to software programing language, which takes time for creation.

### 1.1.3. Material Database

More than six hundred materials are defined in MRADSIM-Converter from National Institute of Standards and Technology (NIST) material data repository [15]. These materials can be assigned to each volume when selected and linked to the GDML file produced. Moreover, new materials with different weight or atomic fractions can be added in case new combinations are required.

### 1.1.4. CAD Processor:

MRADSIM-Converter data import and reading module are performed using Open CASCADE libraries in CAD processor. A new algorithm was implemented to process the geometry structures for reading shapes individually, storing related data e.g., dimensions and position, and maintaining the tree model and hierarchy of the imported STEP file in the assembly structure. The geometry is converted into a detailed mesh structures capable of handling all types of shapes including micro-sized ones. All the related information of data structure is connected to the viewer model where the user interactions are allowed.

## II. MRADSIM-CONVERTER VALIDATION

Validation of MRADSIM-Converter v0.4.4 was carried out to confirm that the volumes and materials are exported correctly to the GDML file after conversion. Its performance was validated in terms of STEP to GDML conversion by two converters, and as an input for simulation in Geant4.

An example geometry of CubeSat with various volumes was used for the validation of MRADSIM-Converter. Fig. 4 shows the imported STEP geometry in MRADSIM-Converter. Two common converters for the GDML format production from CAD files were used for validation: GUIMesh [16], and a commercial software.

Monte Carlo simulations were performed taking into account the GDML files created by MRADSIM-Converter and the other two converters. Geant4 simulation toolkit was used for reading the exported files. All the calculations were done using Dell Inspiron 15 Intel core i7 with 8 threads and 16 GB RAM. In Fig. 5, the GDML geometry imported to Geant4 with the simulation of one thousand geantinos is shown. An isotropic point source of geantino particles was placed in the center of the geometry. Two million events were considered for the simulation. In Table 1, the simulation times is shown, which is the time used for particle transportation for all events. Furthermore, CPU and memory consumption of Geant4 simulation performed for each GDML file produced by the converters is presented. The time for the simulation model using the GDML file produced by MRADSIM-Converter is in a good agreement with other converters, having negligible differences.

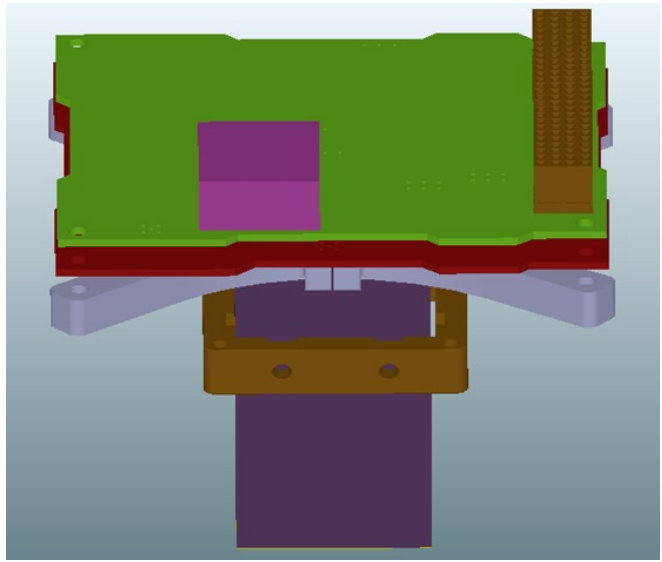

FIG. 4: STEP file imported in MRADSIM-Converter for GDML format conversion

| Converter | No. of events | Time (s) | | | CPU(%) | Memory (MB) | File size (MB) |
|---|---|---|---|---|---|---|---|
| | | User | Real | System | | | |
| GUIMesh | 2000000 | 594.86 | 75.46 | 0.28 | 788.7 | 163 | 9.9 |
| Commercial software | 2000000 | 583.17 | 74.34 | 0.32 | 784.8 | 140 | 5.9 |
| MRADSIM-converter0.4.4 | 2000000 | 604.30 | 76.44 | 0.33 | 790.9 | 151 | 10.8 |

Table 1: GDML geometry performance in Geant4 simulation after STEP to GDML format conversion from different converters

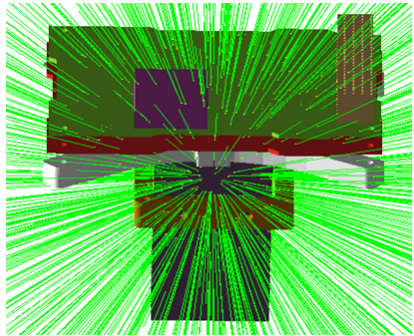

FIG. 5: Geant4 simulation of isotropic geantino source on the converted geometry.

### 3.1. MRADSIM-Converter vs GUIMesh

The advantages of MRADSIM-Converter compared to GUIMesh are as follows.

The GUIMesh code, for converting the STEP to GDML format, requires the user to manually load FREECAD libraries every time the code is executed. Also, for each volume, it creates a GDML file separately, where they are addressed together in another file. This method is time consuming since Geant4 needs to read each file separately. On the other hand, MRADSIM-Converter provides a unique GDML file for the analysis, which reduces the time needed to import the geometry in Geant4.

GUIMesh does not have a 3D visualization tool. Thus, the user cannot visualize the geometry and interact with it through a 3D viewer, although it is possible to assign the material to volumes without visualization. In MRADSIM-Converter, on the contrary, thanks to the user-friendly 3D interface, the user can use both the volumes list and their 3D visualization to interact with the geometry, with the possibility to hide outer volumes to see more internal ones, making it easier to correctly assign materials in complex geometries.

Another feature of MRADSIM-Converter is the possibility of material selection for multi-volumes. This option provides the user a straightforward method for material assignment of various volumes at the same time, which comes to importance when the volumes in the geometry are out-numbered.

### 3.2. MRADSIM-Converter vs a commercial software

The advantages of MRADSIM-Converter compared to a commercial software used in this paper are as follows.

In order to examine the minimum size of a volume acceptable for the STEP to GDML conversion, a test was performed for the conversion of a sphere having the radius of 100 nm. While MRADSIM-Converter and GUIMesh produced the output file for such a tiny volume, the commercial software was unable to create the GDML output format.

The time required for importing a STEP file in MRADSIM-Converter was compared with other converters. A complex geometry of a satellite with the size of 250 MB (shown in Fig. 1) was imported to the converters. The GUIMesh and commercial software took more than 45 minutes for importing the input file while the time needed to MRADSIM-Converter to load the file reduced to 2.5 minutes. A full comparison of the conversion times and output file dimensions between the three converters is given in the following table.

| Converter | Conversion Time | Output file size |
|---|---|---|
| GUIMesh | 55 min | 196 MB |
| Commercial software | 45 min | 80 MB |
| MRADSIM | 2.5 min | 149 MB |

*Table 2: Comparison of the conversion performance of different converters with a 250 MB input STEP file.*

### III. IMPACT

MRADSIM-Converter is a 3D software for CAD to GDML format conversion. Complex geometries can be imported into the MRADSIM-Converter and materials can be assigned to the existing volumes. The converted GDML format is then used for radiation analysis in Geant4-based Monte Carlo codes. To our knowledge this the unique free code with performances better than any software tool available on the market. Using the MRADSIM-Converter provides GEANT4 user to import also very large geometries created by engineers in CAD format. This conversion saves to Geant4 user a large amount of time and operational complexity as well as saving them from the errors that they may introduce by inserting the complex geometries by hand. This in turn, provides engineers fast feedback from GEANT4 simulation and gives opportunity, in a fast turn, to optimize their geometries until the best configuration is obtained.

Indeed, large numbers of users from all around the world have requested a copy of it for use in their projects. These includes but not limited to institutions and agencies such as ESA, NASA, MIT, INFN, CAS, BNL-USA, JLAB-USA, CDRN etc [17]. The MRADSIM project have been a precursor for the creation of an innovative spin-off of INFN, BEAMIDE srl. This spin- off company have then received local (Umbria Region) and National (Italian and Turkiye) funds for development of upcoming versions of the tool.

### IV. CONCLUSION

A Monte Carlo simulation was carried out to validate MRADSIM-Converter performance with two different converters. Its conversion performance is comparable with those tested in this paper, having higher conversion speed with the possibility of converting the tiny shapes to GDML format, and multi-task GUI which allows the complete control for volumes visualization and material assignment in a straightforward way. MRADSIM-Converter is a useful software having user-friendly interfaces permitting the user to import STEP geometries with arbitrary size and complexity and converting them to GDML format. A non-commercial, reduced-functionality version of MRADSIM-Converter is already circulating in the research community and has received remarkable feedbacks, which is very promising for the subsequent development of commercial versions. The ongoing collaboration of researchers from BEAMIDE, IRADETS and INFN has led to the development of a new software called MRADSIM-Space which is going to be released in near future. A complete 3D engineering software for radiation analysis with all the required modules already implemented having the visualization tools for geometry where the user has the possibility to import CAD or GDML files, assign the materials, specify an internal or external source for the simulation, define the required parameters for radiation performance analysis for components, add necessary physics model, and visualize the output results. The main goal of MRADSIM-Space is to combine high computational performance with a simple graphical interface that allows even non-expert programmers to effectively simulate the effects of radiation on devices.

### V. FUTURE PLANS

The future plans includes to release MRADSIM-Converter-Pro which will be a professional version of the Converter with additional features about the modification of the geometries, including and modifying also GDML files as well as the material information reading directly from input files.

The MRADSIM-Space version is ready to be released on the market. This version will include all features of MRADSIM- Converter-Pro and provide user possibility to simulate the radiation effects of components of interest into for instance one or more components located in a large satellite. Following the MRADSIM-Space, there will be a version optimized for Earth based applications (MRADSIM-Earth) such as radioprotection studies of particle accelerators, radiotherapy rooms, nuclear implants etc. The progress of the project will go on to provide Web based versions of all above packages.

### ACKNOWLEDGEMENTS

The authors would like to express their sincere gratitude to the people from the University of Stanford, ESA, CERN, INFN and other users institutes for their time and effort to test and validate this code and for their fruitful comments which improved the performance and quality of MRADSIM-Converter.